\documentstyle[12pt]{article}
\begin{document}
\newcommand{\be}{\begin {equation}}
\newcommand{\ee}{\end{equation}}

\begin{center}

{\bf Is QCD asymptotically free?}\\

\vspace{3mm}

I.M. Dremin\footnote{Email: dremin@lpi.ru}\\

Lebedev Physical Institute, Moscow 119991, Russia\\

\end{center}

\begin{abstract}
The asymptotical behaviour of the QCD coupling strength is considered.
Its upper bound is found from the present experimental data. The assumption
about the finite asymptotical value of the coupling strength leads to the
energy scale where the perturbative predictions would fail. Possible tests
of this assumption are considered.
\end{abstract}

The property of the asymptotical freedom of the QCD coupling strength
$\alpha _S$ is one of the most important ingredients which justifies
the perturbative approach at high momentum transfer. The running $\alpha _S$
has been predicted by the perturbative calculations and confirmed by
experiments at presently available energies. However at higher energies
the dynamics and, consequently, the coupling strength behaviour can change.
If our ideas about the grand unification theory (GUT) are valid, the coupling
strengths of strong, electroweak and gravitation interactions should tend to
a common asymptotic limit.

The fine structure constant of the electromagnetic forces is supposed to
increase with the transferred momentum. The standard perturbative QED
statement implies that it will become infinite in asymptotics if its present
value 1/137 is taken into account. Correspondingly, if the
asymptotic limit is finite, then the perturbative arguments tell us that the
electromagnetic interaction vanishes (the so-called "Moscow zero" effect).
At higher momentum transfers the interacting partners penetrate deeper and
the screening effect of the polarization cloud becomes less pronounced.
The observed value of the charge should be smaller due to the partial screening
by the virtual pairs. Nevertheless, there were proposals \cite{glow} to
consider the variant with the finite limit for the "bare" charge.

Thus, what concerns the limiting value $\alpha _{\infty}$ of all coupling
strengths, it must be, at least, larger than 1/137 and lower than the
$\alpha _S$ values measured at LEP (about 0.118) because the strong coupling
should decrease with the momentum transfer increase. If this tendency
persists asymptotically and there is a common limit with no crossover in the
behaviour of various coupling strengths, then $\alpha _S$ should tend to this
limit from above. This would imply that QCD is not asymptotically free
in a rigorous sense. Let us consider such a possibility.

The traditional lowest order expression for the running QCD coupling strength is
written as
\be
\alpha _S(s)=\frac {4\pi }{\beta _0\ln \frac {s}{Q_0^2}},    \label{alfa}
\ee
where $s$ is the squared momentum transfer (or energy for $e^+e^-$-processes),
$\beta _0=11-\frac {2}{3}n_f$ ($n_f$ is the number of active flavours).
$Q_0$ is a cut-off parameter. It can be chosen (see, e.g., \cite{rady}) as
$Q_0=0.65\Lambda _{\bar {MS}}$ where $\Lambda _{\bar {MS}}$ is the cut-off
of the so-called $\bar {MS}$-renormalization scheme. With such a parameter
the formula (\ref{alfa}) actually reproduces quite well the higher order
expression in  $\bar {MS}$-scheme for $\alpha _S$. In what follows we use
(\ref{alfa}).

The value of $Q_0$ can be determined from (\ref{alfa}) once the mass of
$Z^0$-boson $M_Z\approx 91.2$ GeV and $\alpha _S(M_Z)\approx 0.118$ are known.
One gets
\be
Q_0=M_z\exp \left(-\frac {2\pi }{\beta _0\alpha _S(M_Z)}\right )= 0.246 \; GeV \;\;
(0.153; 0.088).   \label{q0}
\ee
Here and everywhere below the initial number corresponds to $n_f=3$ and the
numbers in brackets to $n_f=4$ and 5.

The formula (\ref{alfa}) describes the asymptotically free behaviour of the
coupling strength $\alpha _S(s)\rightarrow 0$ at $s\rightarrow \infty$.
This would imply the crossover with the electrodynamical fine structure
constant. To avoid such crossover and provide the finite asymptotical limit
for $\alpha _S(s)$ the following natural generalization of the expression
(\ref{alfa}) is proposed:
\be
\alpha _S^{(m)}(s)=\frac {4\pi }{\beta _0\ln \frac {ss_0}{(s+s_0)Q_0^2}}.
\label{alfm}
\ee
The modified coupling strength $\alpha _S^{(m)}(s)$ coincides with the running
coupling strength
(\ref{alfa}) at low energies and tends asymptotically to the finite limit
\be
\alpha _{\infty }=\frac {4\pi }{\beta _0\ln \frac {s_0}{Q_0^2}}. \label{alfi}
\ee
It should be the common limit for all interactions in GUT.
For this generalization one has to pay by the new energy scale parameter $s_0$:
\be
s_0^{1/2}=Q_0\exp \left(\frac {2\pi }{\beta _0\alpha _{\infty }}\right ).  \label{s0}
\ee

Let us consider what corrections to the ordinary $\alpha _S$ are introduced by
this parameter. From (\ref{alfm}) and (\ref{alfa}) one gets
\be
\alpha _S^{(m)}(s)=\alpha _S(s)\left[1-\frac {\ln (1+\frac {s}{s_0})}
{\ln \frac {s}{Q_0^2}}\right ]^{-1}.    \label{amaa}
\ee
In the low energy region $Q_0^2<s\ll s_0$ it reduces to
\be
\alpha _S^{(m)}(s)\approx \alpha _S(s)\left[1-\frac {s}
{s_0\ln \frac {s}{Q_0^2}}\right ]^{-1}.    \label{amal}
\ee
Thus the power-like correction to (\ref{alfa}) is provided by the scale $s_0$.
At very high energies $s\gg s_0$ the coupling strength tends to its
asymptotical limit also in a power-like manner:
\be
\alpha _S^{(m)}(s)\approx \alpha _{\infty }\left[1+\frac {s_0}
{s\ln \frac {s_0}{Q_0^2}}\right ].    \label{amas}
\ee

The upper bound on the constant $\alpha _{\infty }$ is imposed by the
accuracy of the present experimental data on $\alpha _S(M_Z)$:
\be
\alpha _{\infty }\leq \frac {\alpha _S(M_Z)}{1-x\ln (e^{\frac {\delta }{x}}-1)},
\label{aimz}
\ee
where
\be
x=\frac {\beta _0\alpha _S(M_Z)}{4\pi }     \label{x}
\ee
and $\delta $ denotes the relative error in the determination of
$\alpha _S(M_Z)$. For $\delta \ll x$ the good estimate is provided by the
approximate expression:
\be
\alpha _{\infty }\leq \frac {\alpha _S(M_Z)}{1-\frac {\delta }{2} +
x\ln \frac {x}{\delta }}.   \label{aiaz}
\ee

With present accuracy \cite{pdg} about 2$\%$ (i.e., $\delta =0.02$) the
following upper bounds are imposed on $\alpha _{\infty }$:
\be
\alpha _{\infty }\leq 0.1061 \;\;\; (0.1076; 0.1090).  \label{inf}
\ee
It would be necessary to improve the accuracy of the determination of
$\alpha _S(M_Z)$ to the level about 0.03$\%$, to reach the upper bound
on $\alpha _{\infty }$ as low as 0.08. Unfortunately, such accuracy is hardly
achievable at present.

This number is important because there are various arguments \cite{rana, ritu}
in favour of a definite limiting value of $\alpha _{\infty }=1/4\pi
\approx 0.0796$. Actually, these arguments are suggested for the asymptotical
value of the "bare" electromagnetic charge $e_0=\sqrt {\hbar c}$ but the same
limit is valid for strong interactions in GUT. The strong coupling strength is
the closest one to this value (0.118 at $Z^0$) at present. Therefore, it is
most reasonable to try to get some knowledge about $\alpha _{\infty }$ from
QCD-governed processes.

If the limiting value $\alpha _{\infty }=1/4\pi $ is really preferred by nature,
it becomes possible to estimate the characteristic energy scale $s_0$ from
the formula (\ref{s0}):
\be
\sqrt {s_0}\approx 1.6 \; TeV \;\;\; (2; 2.6).   \label{s0e}
\ee
It is seen that this scale is high but, surprisingly enough, much lower than
Planck scales. It lies in the reach of the next generation of colliders.
Nevertheless, to confirm or disprove these scales in experiment is
very difficult if not problematic. To demonstrate this statement, we consider
two methods of $\alpha _S$ determination in $e^+e^-$ experiments.

At the beginning, let us estimate the ratio $\alpha _S^{(m)}(s)/\alpha _S(s)$
at various energies. At the $Z^0$ peak they differ by 0,03 - 0.01 per cents
only. At $\sqrt s =1 $ TeV this difference increases to 2 - 3 $\%$. For energies
$\sqrt s =3\sqrt {s_0}$ it can be as large as 10 -15 $\%$. Thus, in principle,
the experiments in the high energy region can distinguish between the two
possibilities.

Now, consider the ratio $R$ of the hadronic to muon cross sections:
\be
R=R_0(1+\frac {\alpha _S(s)}{\pi}+...).     \label{r}
\ee
The difference in $R$ is $\pi $ times smaller than the above differences.
This implies that extremely precise measurements are needed at very high
energies to discover the limiting value $\alpha _{\infty }=1/4\pi $.

Even more dramatic is the situation with the energy dependence of mean
multiplicities in quark and gluon jets. It is known (for a review see, e.g.,
\cite{dgar}) that these multiplicities increase as $\exp c\sqrt {\ln s}$
for the running coupling strength \cite{dkmt} and like a power $s^{\kappa }$
for the fixed coupling \cite{dhwa}. Surely, the power law takes
over the first one at high energies. If the coupling strength stops its
running at high energy, one would await this effect to show up. It really does,
however, at extremely high and unreachable energies.

It is easy to estimate the corresponding exponents in the DLA expressions for
the multiplicities from the abovecited papers. The running coupling behaviour is
\be
\langle n\rangle _r=A\exp (2c\ln^{1/2}\frac {s}{Q_0^2}).   \label{nr}
\ee
Here $c=(4N_c/\beta _0)^{1/2}$, $N_c=3$ is the number of colours, $A$ is a
common normalization constant, defined by the non-perturbative region.

The fixed coupling theory predicts the power-like growth of mean multiplicity
with energy increase:
\be
\langle n\rangle _f=A\exp (c\ln \frac {s}{Q_0^2}/\ln ^{1/2}\frac {s_0}{Q_0^2}).
\label{nf}
\ee
Herefrom, the energy at which the fixed coupling regime would show up is
\be
\sqrt s = Q_0\left( \frac {s_0}{Q_0^2}\right)^2=Q_0\exp \left(\frac {8\pi }
{\beta _0\alpha _{\infty}}\right).      \label{smu}
\ee
These energies are far outside the particle energy region available in nature if
$\alpha _{\infty }=1/4\pi $. They are about 10$^{(23 - 25)}$ eV.
These estimates show that the results of the jet calculations with the running
coupling strength at present energies are safe and do not depend on its
particular asymptotical value.

The only hope to get in the nearest future $\alpha _{\infty }$ appears if its
value is as "high" as 0.1 by some unknown reasons. Then the improvement by a
factor of 2 in accuracy at $Z^0$ peak would already allow to check this
hypothesis. At the energy 1 TeV, the
modified coupling strength would be larger than the traditional perturbative
value by about 15 - 20 $\%$. In addition, some other data (shape variables etc)
could be more sensitive to the variations of the coupling strength. More
complete analysis is needed to compare various upper bounds.

To conclude, the absence of the asymptotical freedom in QCD is hard to confirm
or disprove in experiment in the nearest future if the limiting value of the
coupling strength is $1/4\pi $ or lower. If the asymptotical values of the
coupling strength are very close to 0.1, they can be quantitatively established
when the accuracy of the data processing increases. The low values of the energy
scale parameter $s_0$ are encouraging for new prospects at the next generation
of accelerators. Apart from the considered data, other experimentally measured
characteristics (like event shapes) could be more sensitive to these features of
QCD. The complete analysis of all the data from this point of view should be
done and upper bounds on $\alpha _{\infty }$ determined and compared to answer
the question posed to the title of the paper.

This work has been supported in parts by the RFBR grants N 02-02-16779,
03-02-16134, NSH-1936.2003.2.

\end{document}